# Nearly Tight Low Stretch Spanning Trees

Ittai Abraham[*] Yair Bartal[†] Ofer Neiman[‡]

October 23, 2018

**Abstract**

We prove that any graph $G$ with $n$ points has a distribution $\mathcal{T}$ over spanning trees such that for any edge $(u,v)$ the expected stretch $E_{T\sim\mathcal{T}}[d_T(u,v)/d_G(u,v)]$ is bounded by $\tilde{O}(\log n)$. Our result is obtained via a new approach of building "highways" between portals and a new strong diameter probabilistic decomposition theorem.

## 1 Introduction

Let $G=(V,E)$ be a finite graph. For any subgraph $H=(V',E')$ of $G$ let $d_H$ be the induced shortest path metric with respect to $H$. In particular, for any edge $(u,v)\in E$ and any spanning tree $T$ of $G$, $d_T(u,v)$ denotes the shortest path distance between $u$ and $v$ in $T$.

Given a distribution $\mathcal{T}$ over spanning trees of $G$, let $\mathrm{stretch}_{\mathcal{T}}(u,v)=\mathbb{E}_{T\sim\mathcal{T}}\left[\frac{d_T(u,v)}{d_G(u,v)}\right]$ and let $\mathrm{stretch}_{\mathcal{T}}(G)=\max_{(u,v)\in E}\mathrm{stretch}_{\mathcal{T}}(u,v)$. Let $\mathrm{stretch}(n)=\max_{G=(V,E)||V|=n}\inf_{\mathcal{T}}\{\mathrm{stretch}_{\mathcal{T}}(G)\}$.

Initial results were obtained by Alon, Karp, Peleg and West [2] showing that $\Omega(\log n)=\mathrm{stretch}(n)=\exp(O(\sqrt{\log n\log\log n}))$. The upper bound was significantly improved to $O((\log n)^2\log\log n)$ by Elkin, Emek, Spielman and Teng [10][1]. For the class of Series-Parallel graphs Emek and Peleg [11] obtained a bound of $\Theta(\log n)$. The main result of this paper is a new upper bound on $\mathrm{stretch}(n)$ that is tight up to polylogarithmic factors[2].

**Theorem 1.**

$$\mathrm{stretch}(n)=O\left(\log n\cdot\log\log n\cdot(\log\log\log n)^3\right)$$

**Remark 1.** *For ease of presentation we first show a slightly weaker bound of*

$$\mathrm{stretch}(n)=O\left(\log n\cdot(\log\log n)^2\cdot\log\log\log n\right),$$

*and prove the tighter bound in Appendix B*

Our result may be applied to improve the running time of the Spielman and Teng [16] solver for sparse symmetric diagonally dominant linear systems.

[*]School of Engineering and Computer Science, Hebrew University, Israel. Email: ittaia@cs.huji.ac.il.

[†]School of Engineering and Computer Science, Hebrew University, Israel and Center of the Mathematics of Information, Caltech, CA, USA. Email: yair@cs.huji.ac.il. Supported in part by a grant from the Israeli Science Foundation (195/02) and in part by a grant from the National Science Foundation (NSF CCF-065253).

[‡]School of Engineering and Computer Science, Hebrew University, Israel. Email: neiman@cs.huji.ac.il. Supported in part by a grant from the Israeli Science Foundation (195/02).

[1]In fact these result apply to a similar notion, $\mathrm{avg}-\mathrm{stretch}(n)=\max_{G=(V,E)||V|=n}\inf_T\{\frac{1}{|E|}\sum_{(u,v)\in E}\frac{d_T(u,v)}{d_G(u,v)}\}$ which is equivalent up to a constant factor to $\mathrm{stretch}(n)$.

[2][9] announced $\mathrm{stretch}(n)=O((\log n)^2)$, but this claim was subsequently withdrawn by the authors

## 1.1 Techniques

We extend the star-decomposition technique of Elkin *et. al.*[10]. A star-decomposition of a graph is a partition of the vertices into clusters that are connected into a star: a central cluster is connected to every other cluster by a single edge. As in [10] given a subgraph over a cluster $X$, the central cluster $X_0$ is formed by cutting a ball with radius $r_0$ around a center $x_0$ and the remaining clusters $X_1, X_2, \ldots,$ which are called cones, are formed iteratively. Let $Y_j = X \setminus \bigcup_{0 \le k \le j} X_k$. The cone $X_j$ is created by choosing an edge $(y_j, x_j)$ such that $y_j \in X_0, x_j \in Y_{j-1}$ and defining $X_j$ as the cone with radius $r_j$ around $x_j$ from the cluster $Y_{j-1}$, as all the points whose distance to $x_0$ going through the edge $(x_j, y_j)$ does not increase too much relatively to the shortest path distance, formally $X_j = \{x \in Y_{j-1} \mid d_X(x_0, y_j) + d_X(y_j, x_j) + d_{Y_{j-1}}(x_j, x) - d_X(x_0, x) \le r_j\}$. Let $\mathrm{rad}_{x_0}(X) = \max_{x \in X} d(x_0, x)$, then typically the radius of the central ball is chosen so that $r_0 \approx \mathrm{rad}_{x_0}(X)/c$ for a constant $c$. An important parameter of a star-decomposition is the radius of the cone. We say that the star-decomposition has parameter $\epsilon$ if for any $j \ge 1$, the radius $r_j$ of the cone $X_j$ is at most $\epsilon \cdot \mathrm{rad}_{x_0}(X)$.

Applying star-decompositions in a recursive manner induces a spanning tree $T$. For a point $u$ denote by $X^{(i)}$ the cluster that contains $u$ in the $i$th recursive invocation of the hierarchical star-decomposition algorithm.

The $O(\log^2 n \log\log n)$ bound of [10] is obtained by choosing $\epsilon \approx 1/\log n$ and showing:

1. $O(1)$ *radius stretch*. For any cluster $X$ induced by the recursive invocation of the hierarchical star-decomposition algorithm, and any $z \in X$, $d_T(x_0, z) = O(\mathrm{rad}_{x_0}(X))$.

2. $O((\log n \cdot \log\log n)/\epsilon)$ *decomposition stretch*. For any edge $(u, v)$, $\sum_i \Pr[(u, v)$ is separated when star-decomposing $X^{(i)}] \cdot \mathrm{diam}(X^{(i)}) = O(\log n \log\log n)/\epsilon$.

Combining these two properties yields their result, noticing that if the end points of an edge $(u, v)$ fall into different clusters in the partitioning of $X^{(i)}$ then $d_T(u, v)$ can be bounded by $d_T(u, x_0) + d_T(v, x_0) = O(\mathrm{diam}(X^{(i)}))$.

Good radius stretch is obtained by observing that in each recursive application of the star partition the radius of a cluster is stretched by at most $1 + 1/\log n$, and since there are $O(\log n)$ scales the total radius stretch is a constant. Good decomposition stretch is obtained by using a version of the decomposition of [4, 8].

**Better radius stretch.** In our scheme we perform a star-decomposition with a parameter $\epsilon \approx 1/\log\log n$, this significantly improves the decomposition stretch, by a factor of $\approx \log n/\log\log n$. A naive attempt to bound the radius stretch, by $1 + 1/\log\log n$ in each scale, will result in super logarithmic radius stretch over all scales.

We introduce a new approach to bound the radius stretch. We arrange all the points of $X$ in a queue $Q = (z_1, z_2, \ldots, z_n)$, and bound the distance $d_T(x_0, z_i)$ as a function of $i$ by building "highways" – *low stretch paths*. Roughly speaking, we obtain a bound of $d_T(x_0, z_i) = O(\log\log i) \cdot \mathrm{rad}_{x_0}(X)$. The core observation is that by choosing where to build the first cone and passing this information into the recursion, one can obtain a shortest path "highway" between $x_0$ and the first point $z_1$, such that the distance between $x_0$ and $z_1$ in the tree will be *exactly* the original distance in the graph. The challenge is to use this observation to maintain "highways" – low stretch paths – between $x_0$ and *all* the points. Specifically, we obtain

1. $O(\log\log n)$ *radius stretch*. For any cluster $X$, and any $z \in X$, $d_T(x_0, z) = O(\log\log n)\mathrm{rad}_{x_0}(X)$.

**Better decomposition stretch.** A relaxation of the spanning tree problem suggested by Bartal [3] is to consider a distribution of dominating tree metrics (in fact of ultrametrics) that do not necessarily span the graph. This relaxation has proven applicable for approximation algorithms, online problems and has contributed to recent solutions for the spanning tree problem (*i.e.* [10]). Initially $O(\log^2 n)$ approximation was obtained in [3] based on the truncated exponential distribution approach of [14]. This bounded was subsequently improved to $O(\log n \log\log n)$ in [4] and [8]. Finally an optimal $O(\log n)$ approximation was obtained by [12] based on the cutting scheme of [7]. Subsequently an $O(\log n)$ bound was also obtained using a truncated exponential distribution approach [5, 1].

However, all previous schemes that obtained the optimal $O(\log n)$ bound for the metric problem were insufficient for the spanning tree problem. Given a graph $G = (X, E)$, a sequence $x_1, x_2, \ldots$ of cluster centers and a sequence $r_1, r_2, \ldots$ of radiuses we can define a weak diameter decomposition by defining $W_i = B_X(x_i, r_i) \setminus \bigcup_{j<i} W_j$. We can define a strong diameter decomposition by defining $C_i = B_{X \setminus \bigcup_{j<i} C_j}(x_i, r_i)$. Observe that in a strong diameter decomposition, for any nonempty cluster $C_i$, we have that $x_i \in C_i$ and $C_i$ is a connected component of $G$, this may not be the case for weak diameter decompositions. Indeed the techniques of [12, 5, 1] provide a weak diameter decomposition. It was not clear how to extend these results to strong diameter decompositions that are necessary for star-decompositions. We show how to obtain a strong diameter hierarchical decomposition theorem that obtains an optimal bound in the following sense:

2. $O(\log n \log(1/\epsilon)/\epsilon)$ *decomposition stretch.* For any edge $(u, v)$, $\sum_i \Pr[(u, v)$ is separated when star-decomposing $X^{(i)}] \cdot \operatorname{diam}(X^{(i)}) = O(\log n \log(1/\epsilon)/\epsilon)$.

As in [5, 1], our decomposition is based on the truncated exponential distribution with a parameter depending on the local growth rate of the space. The main technical difficulty arises since the space *changes* after each cluster is cut (the metric is derived from a graph, and some nodes and edges are removed at every cut). The idea is to define the local growth rate with respect to the current metric, and to show two things: that the expected sum of all growth rates (which are random variables) over all the scales telescopes to $n$, and that the probability to be cut is appropriately bounded in each scale. Dealing with the randomly changing graph raises some additional subtleties in the proof. Our strong diameter hierarchical decomposition theorem may be of independent interest.

### 1.2 Applications

One of the main applications of low stretch spanning trees is solving *sparse symmetric diagonally dominant linear systems of equations*. This approach was suggested by Boman and Hendrickson [6] and later improved by Spielman and Teng [16]. Spielman and Teng showed an algorithm that for such an $n$-by-$n$ matrix $A$ with $m$ non-zero entries and an $n$-dimensional vector $b$, if $\epsilon > 0$ is the precision of the solution then the algorithm finds $x'$ such that $\|x - x'\|_A \le \epsilon$ where $Ax = b$, and the running time is $O\left(m\left(\log^{O(1)} m + \log(1/\epsilon)\right) + n \cdot \mathrm{avg} - \mathrm{stretch}(n) \cdot \log(1/\epsilon)\right)$. Improving the bound requires improvement of the second element, and we improve it by roughly an additional $O(\log\log n)$ factor over [10]. Actually, if the running time of our construction is reduced, we can obtain an $O(\log n)$ improvement. For planar graphs we obtain $O(n \cdot \log^2 n)$. See details in Corollary 6.

The *minimum communication cost spanning tree* problem introduced in [13], in which one is given a weighted graph $G = (V, E, w)$ and a matrix $A = a_{xy} \mid x, y \in V$, the objective is to find a spanning tree minimizing $c(T) = \sum_{x,y\in V} a_{xy} \cdot d_T(x, y)$. [15] showed an $O(2^{\sqrt{\log n \cdot \log\log n}})$ approximation ratio based on [2], and [10] improved to $O(\log^2 n \cdot \log\log n)$. Our results can be used to obtain $O(\log n \cdot \log\log n(\log\log\log n)^3)$ approximation ratio.

See [10] for details about more applications.

### 1.3 Structure of the Paper

In Section 2 we describe a star-decomposition framework, that for any unweighted $n$ point graph $G$ induces a tree such that $\operatorname{diam}(T) \le O(\operatorname{diam}(G) \cdot \log\log n)$. In Section 3 we describe a distribution on star-partitions that follows the framework of Section 2. We analyze the expected stretch of an edge and prove the bound of $\operatorname{stretch}(n) = O(\left(\log n \cdot (\log\log n)^2 \cdot \log\log\log n\right))$. In Appendix A we discuss briefly how to extend the result for weighted graphs. In Appendix B we show the tighter result stated in Theorem 1.

## 2 Highways

Let $G = (V, E)$ be a finite graph. For any $X \subseteq V$ let $d_X : X^2 \to \mathbb{R}^+$ be the shortest path metric induced by the subgraph on $X$. Let $\operatorname{diam}(X) = \max_{y,z\in X}\{d_X(y, z)\}$. For $x \in X$ let $\operatorname{rad}_x(X) = \max_{y\in X} d_X(x, y)$, we omit the subscript when clear from context (note that $\operatorname{diam}(X)/2 \le \operatorname{rad}(X) \le \operatorname{diam}(X)$). For any $x \in X$ and $r \ge 0$ let $B_{X,d}(x, r) = \{y \in X \mid d_X(x, y) \le r\}$. Let $c = 2^{16}$ be a constant. We use the uppercase letter $Q$ to denote a *queue*, a sequence of points. Given a point $x$ not in the queue we say that we enqueue $x$ into $Q$ meaning that we add $x$ as the last element of the sequence and given a queue $Q$, the dequeue operation removes and returns the first element of the sequence.

**Definition 1** (cone metric[3]). *Given a graph $G = (V, E)$, subsets $Y \subset X \subseteq V$, points $x \in X \setminus Y$ , $y \in Y$ define the* cone-metric $\rho = \rho(X, Y, x, y) : Y^2 \to \mathbb{R}^+$ *as* $\rho(u, v) = |(d_X(x, u) - d_Y(y, u)) - (d_X(x, v) - d_Y(y, v))|$.

Note that a ball $B_{Y,\rho}(y, r)$ in the cone-metric $\rho = \rho(X, Y, x, y)$ is the set of all points $z \in Y$ such that $d_X(x, y) + d_Y(y, z) - d_X(x, z) \leq r$.

**Hierarchical-Star-Partition algorithm.** See Figure 1 for the algorithm. Given an unweighted graph $G = (V, E)$, create a spanning tree $T = (V, E')$ by choosing some $x_0 \in V$, letting $Q$ be an arbitrary ordering of $V \setminus \{x_0\}$ and calling: `hierarchical-star-partition`$(V, x_0, Q)$.

$T =$ `hierarchical-star-partition`$(X, x_0, Q)$:

1. If $\text{rad}_{x_0}(X) \leq 16c$ return BFS$(X)$.
2. $(X_0, \dots, X_m, (y_1, x_1), \dots, (y_m, x_m), Q_0, Q_1, \dots, Q_m) =$ `star-partition`$(X, x_0, Q)$;
3. For each $i \in [0, \dots, m]$:
4. $T_i =$ `hierarchical-star-partition`$(X_i, x_i, Q_i)$;
5. Let $T$ be the tree formed by connecting $T_0$ with $T_i$ using edge $(y_i, x_i)$ for each $i \in [1, \dots, m]$;

Figure 1: `hierarchical-star-partition` algorithm

**Star-Partition algorithm.** See Figure 2 for our `star-partition` algorithm. We highlight the main differences of our algorithm from that of [10]. In addition to $X, x_0$ it receives as input an ordering of the points in $X$, implemented as a queue data structure and denoted by $Q$. In addition to returning a star decomposition $X_0, X_1, \dots, X_m$ it returns for each $0 \leq j \leq m$ an ordering of the points in $X_j$, implemented as a queue data structure and denoted by $Q_j$.

Since as noted above the trivial radius bound (loosing $(1 + \epsilon)$ in every scale) does not work anymore we attempt to directly bound $d_T(x_0, z)$ for all $z \in X$. The arrangement of $X \setminus \{x_0\}$ in a queue $Q = (z_1, \dots, z_{n-1})$ determines "how hard" we try to give a tight bound for the point $z_i$ - roughly speaking the smaller value of $i$ means the harder we try to give a better bound on $d_T(x_0, z_i)$. The star partition algorithm therefore changes to try hardest for the first point $z_1$, and indeed by choosing the first portal edge $(y_1, x_1)$ on a shortest path to $z_1$ and keeping $z_1$, $y_1$ in the head of the recursive queues we obtain a "highway" from $x_0$ to $z_1$, *i.e.* preserving the original distance. Surprisingly, this small change is enough to give a good bound on $d_T(x_0, z_i)$ for all $i > 1$, and we obtain $d_T(x_0, z_i) = O(\log\log i)\text{rad}_{x_0}(X)$. The intuition is that since every cluster contains less points, $z_i$ advances in the recursive queues, and when it becomes the first we get a "highway" to it. For this intuition to work one must delicately define the ordering of the queues $Q_0, \dots, Q_m$ for the clusters $X_0, \dots, X_m$ created by the star partition algorithm. The main difficulty is defining $Q_0$, as the portals $y_j$ play a dual part - we need to maintain their original position in $Q$ and also make sure that the tree distance to them is small enough: as it determines the distance from $x_0$ to all the points in $X_j$.

Suppose $z_i \in Y_j$ for some $i > 1$. By Claim 2 there is an inherent loss of a $1+\epsilon$ factor due to star-partition algorithm. Hence it is not sufficient for the inductive argument to simply obtain a bound of $d_T(x_0, y_j) = O(\log\log i)\text{rad}_{x_0}(X_0)$ in the ball $X_0$ and $d_T(x_j, z_i) = O(\log\log i)\text{rad}_{x_j}(X_j)$ in the cone $X_j$. We must "gain" inductively either in $d_T(x_0, y_j)$ (the ball part of the path) or in $d_T(x_j, z_i)$ (the cone part of the path). This is done by choosing the queues in the following manner: Given a star decomposition $X_0, X_1, \dots, X_m$ we create the queue $Q_j$ for $j > 0$ simply as the restriction of $Q$ on $X_j \setminus \{x_j\}$. The queue $Q_0$ is the created by first adding either $z_1$ or the portal $y_1$ which is chosen on a shortest path to $z_1$, thus making sure the distance from $x_0$ to $z_1$ is preserved in the recursion. Then interleaving three different queues $Q_0^{(\text{ball})}, Q_0^{(\text{fat})}, Q_0^{(\text{reg})}$.

- $Q_0^{(\text{ball})}$ is the restriction of $Q$ on $X_0$. This queue provides the required bound on $d_T(x_0, z_i)$ when $z_i \in X_0$.
- $Q_0^{(\text{reg})}$ is a queue of portals $y_j$ ordered by the minimal point of $Q$ that their cones $X_j$ contains. When a cone contains relatively few point we "gain" in the cone part of the path to $z_i$. This queue guarantees that for any $z_i \in X_j$ the "central ball" part of the path to $z_i$ is not stretched too much.

[3] In fact, the cone-metric is a pseudo-metric.

- $Q_0^{\text{(fat)}}$ is a queue of portals $y_j$ that lead to cones that contain "many" points relative to the ordering $Q$ of the points in $X_j$. When a cone is "fat" we cannot gain in the cone part, this queue guarantees that we gain in the ball part.

The exact way these three queues are created is detailed in Line 5 of Figure 2.

$(X_0, \ldots, X_m, (y_1, x_1), \ldots, (y_m, x_m), Q_0, Q_1, \ldots, Q_m) = \texttt{star-partition}(X, x_0, Q)$:

1. Let $j = 2$; Denote the (ordered) elements of $Q$ by $Q = (z_1, z_2, \ldots, z_k)$; Let $\epsilon = \epsilon(X) \in (0, \frac{1}{170c}]$;
2. Creating the ball $X_0$:
   (a) Choose $r_0$ uniformly at random from the interval $[1/(16c), 1/(8c)]$;
   (b) Let $X_0 = B(x_0, r_0 \cdot \text{rad}_{x_0}(X))$; Let $Y_0 = X \setminus X_0$;
3. Creating the first cone $X_1$:
   (a) If $z_1 \in Y_0$ let $z = z_1$ otherwise let $z \in Y_0$ be an arbitrary point. Let $(y_1, x_1)$ be an edge such that $y_1 \in X_0$, $x_1 \in Y_0$ and $d_X(x_0, z) = d_X(x_0, y_1) + d_X(y_1, x_1) + d_{Y_0}(x_1, z)$ (*i.e.* an edge on a shortest path from $x_0$ to $z$);
   (b) Let $\rho = \rho(X, Y_0, x_0, x_1)$ be the cone-metric;
   (c) Choose $r_1$ uniformly at random from the interval $[\epsilon/4, \epsilon/2]$;
   (d) Let $X_1 = B_{(Y_0,\rho)}(x_1, r_1 \cdot \text{rad}_{x_0}(X))$; Let $Y_1 = Y_0 \setminus X_1$;
4. Creating the remaining cones $X_2, \ldots, X_m$:
   (a) While $Y_{j-1} \neq \emptyset$ :
      i. Let $(x_j, y_j, r_j) = \texttt{cone-cut}(X, x_0, X_0, Y_{j-1}, \epsilon)$; (has the property that $r_j \leq \epsilon/2$)
      ii. Let $\rho = \rho(Y_{j-1} \cup X_0, Y_{j-1}, x_0, x_j)$;
      iii. Let $X_j = B_{(Y_{j-1},\rho)}(x_j, r_j \cdot \text{rad}_{x_0}(X))$; $Y_j = Y_{j-1} \setminus X_j$;
      iv. Let $j = j + 1$;
5. Creating the queues $Q_0^{\text{(ball)}}, Q_0^{\text{(fat)}}, Q_0^{\text{(reg)}}, Q_1, \ldots, Q_m$:
   (a) For $i = 1, \ldots, |X| - 1$ :
      i. If $z_i \in X_0$ then enqueue $z_i$ into $Q_0^{\text{(ball)}}$;
      ii. Otherwise let $\ell \geq 1$ be such that $z_i \in X_\ell$:
         - If $z_i \neq x_\ell$ then enqueue $z_i$ into $Q_\ell$.
         - If $y_\ell \notin Q_0^{\text{(reg)}}$ then enqueue $y_\ell$ into $Q_0^{\text{(reg)}}$.
         - If $|X_\ell \cap \{z_1, \ldots, z_i\}| > \sqrt{i}$ and $y_\ell \notin Q_0^{\text{(fat)}}$ then enqueue $y_\ell$ into $Q_0^{\text{(fat)}}$.
6. Creating the queue $Q_0$:
   (a) Denote $Q_0^{\text{(ball)}} = z_1^1, \ldots, z_{m_1}^1$, $Q_0^{\text{(fat)}} = z_1^2, \ldots, z_{m_2}^2$, $Q_0^{\text{(reg)}} = z_1^3, \ldots, z_{m_3}^3$.
   (b) Create $Q_0$ by interleaving the three queues $Q_0^{\text{(ball)}}, Q_0^{\text{(fat)}}, Q_0^{\text{(reg)}}$ such that:
      - If $z_1 \in X_0$ then $z_1$ is the first element of $Q_0$. Otherwise $y_1$ is the first element of $Q_0$.
      - For any $x \in X$, $\ell \in \{1, 2, 3\}$, $1 \leq i \leq n$ if $x = z_i^\ell$ then $x$ is in the first $3i$ elements of $Q_0$.

Figure 2: `star-partition` algorithm

## 2.1 Bounding the radius stretch

In this part we show that the radius stretch induced by the `hierarchical-star-partition` algorithm is at most $O(\log \log n)$.

The following two claims imply that the `star-partition` algorithm on a cluster $X$ induces a partition on $X$ and that radial distances are stretched by a most $1+\epsilon$. These claims are essentially proven in [10] we provide a proof for completeness.

**Claim 1.** *For any graph $X$, $x_0 \in X$, $j > 0$ let $Y_{j-1} \subseteq X$ be the unassigned points of $X$ after creating $j$ clusters $X_0, \dots, X_{j-1}$ using the `star-partition` algorithm, then for any $z \in Y_{j-1}$ all the shortest paths from $z$ to $x_0$ are fully contained in $Y_{j-1} \cup X_0$, in particular*

$$d_{Y_{j-1} \cup X_0}(x_0, z) = d_X(x_0, z).$$

*Proof.* Let $\Delta = \text{rad}_{x_0}(X)$. Let $P_{z,x_0}$ be a shortest path and assume by contradiction that $P_{z,x_0} \not\subseteq Y_{j-1} \cup X_0$, so let $1 \le i \le j-1$ be the minimal $i$ such that there exists $u \in P_{z,x_0}$ and $u \in X_i$. Let $x_i$ be the portal to the cone $X_i$. By Definition 1 since $u \in X_i$ it must be that in the metric $d' = d_{X_0 \cup Y_{i-1}}$

$$d'(u, x_0) + r_i \cdot \Delta \ge d'(u, x_i) + d'(x_i, x_0).$$

Since $u$ lies on a shortest path from $z$ to $x_0$, the minimality of $i$ suggests that this shortest path is fully contained in $Y_{i-1} \cup X_0$ thus $d'(z, x_0) = d'(z, u) + d'(u, x_0)$, and conclude that

$$d'(z, x_0) + r_i \cdot \Delta = d'(z, u) + d'(u, x_0) + r_i \cdot \Delta \ge d'(z, u) + d'(u, x_i) + d'(x_i, x_0) \ge d'(z, x_i) + d'(x_i, x_0),$$

hence $z$ should be in $X_i$, contradiction. $\square$

**Claim 2.** *Let $(X_0, \dots, X_m, (y_1, x_1), \dots, (y_m, x_m), Q_0, Q_1, \dots, Q_m) =$ `star-partition`$(X, x_0, Q)$ then for any $1 \le j \le m$*

$$\text{rad}_{x_0}(X_0) + d(y_j, x_j) + \text{rad}_{x_j}(X_j) \le (1+\epsilon)\text{rad}_{x_0}(X),$$

*Proof.* Let $\Delta = \text{rad}_{x_0}(X)$. Let $\beta$ be such that $\text{rad}_{x_0}(X_0) = \beta \cdot \Delta$, let $d' = d_{X_0 \cup Y_{j-1}}$, let $x_j$ be the portal of $X_j$ and $\rho = \rho(X_0 \cup Y_{j-1}, Y_{j-1}, x_0, x_j)$ be the cone-metric. Take $z \in X_j$ as the farthest point from $x_j$ (with respect to $d'$), take any shortest path $P_{x_j,z}$ from $x_j$ to $z$ and separate it into consecutive segments $x_j = u_0, v_0, u_1, v_1, \dots, u_k, v_k = z$ such that for any $0 \le i \le k$, $\rho(u_i, v_i) = 0$, *i.e.*

$$d'(x_0, u_i) - d'(x_j, u_i) = d'(x_0, v_i) - d'(x_j, v_i)$$

and $(v_i, u_{i+1}) \in E$ (note that it could be that $u_i = v_i$). The definition of cone-metric suggests that $k \le r_i \cdot \Delta$, as otherwise $z \notin B_{Y_{j-1},\rho}(x_j, r_j \cdot \Delta) = X_j$.

Since $P_{x_j,z}$ is a shortest path we have for all $0 \le i \le k$ that $d'(x_j, u_i) + d'(u_i, v_i) = d'(x_j, v_i)$, therefore

$$\sum_{i=0}^{k} d'(x_0, v_i) = \sum_{i=0}^{k} (d'(x_0, u_i) + d'(u_i, v_i)). \tag{1}$$

Claim 1 suggests that $d_X(x_0, z) = d'(x_0, z)$, hence

$$\begin{aligned}
\Delta &\ge d_X(x_0, z) = d'(x_0, z) = d'(x_0, v_k) \\
&= \sum_{i=0}^{k-1} (d'(x_0, u_i) + d'(u_i, v_i) - d(x_0, v_i)) + d'(x_0, u_k) + d'(u_k, v_k) \\
&\ge \sum_{i=0}^{k-1} (d'(x_0, u_i) + d'(u_i, v_i) - (d'(x_0, u_{i+1}) + d'(v_i, u_{i+1}))) + d'(x_0, u_k) + d'(u_k, v_k) \\
&= d'(x_0, u_0) - d'(x_0, u_k) + \sum_{i=0}^{k-1} (d'(u_i, v_i) - 1) + d'(x_0, u_k) + d'(u_k, v_k) \\
&= (\beta\Delta + 1) - k + \sum_{i=0}^{k} d'(u_i, v_i)
\end{aligned}$$

The second line follows from (1), the third from the fact that $d'(x_0, v_i) \leq d'(x_0, u_{i+1}) + d'(u_{i+1}, v_i)$, the fourth since the sum telescopes and $d'(v_i, u_{i+1}) = 1$, and the fifth since $d'(x_0, u_0) = d'(x_0, x_j) = d'(x_0, y_j) + d'(y_j, x_j) = \mathrm{rad}_{x_0}(X_0) + 1 = \beta\Delta + 1$.

Therefore

$$\mathrm{rad}_{x_j}(X_j) = d'(x_j, z) = \sum_{i=0}^{k} d'(u_i, v_i) + \sum_{i=0}^{k-1} d'(v_i, u_{i+1}) \leq (\Delta - \beta\Delta + k - 1) + k \leq (1-\beta)\Delta + 2r_j\Delta - 1,$$

(recall that $k \leq r_j\Delta$). And now since $r_j \leq \epsilon/2$,

$$\mathrm{rad}_{x_0}(X_0) + d(y_j, x_j) + \mathrm{rad}_{x_j}(X_j) \leq \beta\Delta + 1 + (1-\beta)\Delta + \epsilon\Delta - 1 = (1+\epsilon)\Delta.$$

□

**Corollary 3.** *For any* $0 \leq j \leq m$, $\mathrm{rad}_{x_j}(X_j) < (1 - \frac{1}{20c})\mathrm{rad}_{x_0}(X)$

*Proof.* The corollary is immediate for $X_0$ by the construction, for $j > 0$: as $\mathrm{rad}_{x_0}(X_0) \geq \mathrm{rad}_{x_0}(X)/(16c)$ and $\epsilon \leq 1/(170c)$ using Claim 2

$$\mathrm{rad}_{x_j}(X_j) < (1+\epsilon)\mathrm{rad}_{x_0}(X) - \mathrm{rad}_{x_0}(X_0) \leq (1 - 1/(20c))\mathrm{rad}_{x_0}(X).$$

□

**Lemma 4.** *Let $X \subseteq V$ be a connected component of $G(V,E)$. Let $x_0 \in X$ and $Q = (z_1, \ldots, z_{|X|-1})$ be any ordering of $X \backslash \{x_0\}$. Let $T$ be any spanning tree of $G$ returned by the algorithm* `hierarchical-star-partition`$(X, x_0, Q)$ *with parameter $\epsilon = \epsilon(X) = \frac{1}{170c \log\log(|X|)}$, then*

$$d_T(x_0, z_i) \leq \begin{cases} d_X(x_0, z_i) & i = 1 \\ i \cdot \mathrm{rad}_{x_0}(X) & 1 < i < c \\ c \cdot \log\log i \cdot \mathrm{rad}_{x_0}(X) & \textit{otherwise} \end{cases}$$

*(where $c = 2^{16}$)*

*Proof.* The proof is by induction on the radius of $X$. In the base case when $\mathrm{rad}_{x_0}(X) \leq 16c$ create a breadth first tree centered in $x_0$, and since in such a tree for every $z \in X$, $d_X(x_0, z) = d_T(x_0, z)$ the claim holds. Now we turn to the inductive step. Note that Corollary 3 guarantees that for all $j = 0, \ldots, m$ we have $0 \leq \mathrm{rad}_{x_j}(X_j) < \mathrm{rad}_{x_0}(X)$.

The main idea of the proof is to consider a single application of the star-partition algorithm, partitioning $X$ into $X_0, X_1, \ldots, X_m$. Assuming that $z_i \in X_j$ the path between $x_0$ to $z_i$ will be the path going through the edge $(y_j, x_j)$. Then use the induction hypothesis on the sub-path $x_0, y_j$ in $X_0$ and the sub path $x_j, z_i$ in $X_j$. Since by Claim 2 the radius may increase by a factor of at most $1 + \epsilon$, we need to "gain" in one of the two sub paths. This "gain" will occur since our construction guarantees that either the position of $z_i$ in the queue of $X_j$ will improve or the position of $y_j$ in $X_0$ will improve, thus the induction hypothesis will give the required bounds.

There are three main cases to consider, when $i = 1$, $i < c$ and $i \geq c$. The case $i = 1$ is simple. The case $1 < i < c$ subdivides into three more cases:

1. The first case is $z_i \in X_0$. This case is relatively straightforward.

2. The second case is that the first $i$ points of the queue are all in $X_1$. Here we gain in the central ball because the portal $y_1$ leading to $X_1$ will be the first element in $Q_0$.

3. The remaining case is that not all of the first $i$ points are in $X_1$, then there are at most $i - 1$ points in the cone $X_j$ among $z_1, \ldots, z_i$, so by the construction of $Q_j$, we gain just enough in the cone (because the bound that needs to be shown is weak - linear in $i$) and $Q_0^{(\mathrm{reg})}$ guarantees that we do not lose too much in the central ball.

The interesting case is when $i \geq c$, this last case also subdivides into three more cases:

1. One first is that $z_i \in X_0$. Again. this case is relatively straightforward and uses the construction of $Q_0^{(\text{ball})}$.
2. The second case is that $z_i \in X_j$ and $X_j$ is a "thin" cone - contains less than $\sqrt{i}$ of the first $i$ points. Here we gain in the cone because the position of $z_i$ in $Q_j$ is at most $\sqrt{i}$ and $Q_0^{(\text{reg})}$ guarantees that we do not lose too much in the central ball.
3. The third case is that $z_i \in X_j$ and $X_j$ is a "fat" cone - contains more than $\sqrt{i}$ of the first $i$ points. Here we gain in the central ball, using the construction of $Q_0^{(\text{fat})}$ and Claim 5 to show that the portal $y_j$ leading to the cone is in position $\leq i^{9/10}$ in $Q_0$.

We continue with the formal proof of the lemma, according to the three main cases.

**Case 1:** In this case $i = 1$. Note that $z_1 \in X_0 \cup X_1$. If $z_1 \in X_0$ then by the construction $z_1$ is going to be the first in $Q_0$ therefore by the induction hypothesis on $X_0$ it follows that $d_T(x_0, z_1) \leq d_X(x_0, z_1)$. If on the other hand $z_1 \in X_1$, then again from the construction the point $y_1$, which was chosen such that $y_1, x_1$ are on a shortest path from $x_0$ to $z_1$, will be the first in $Q_0$, and $z_1$ will be the first in $X_1$, so by induction $d_T(x_0, z_1) = d_T(x_0, y_1) + d_T(y_1, x_1) + d_T(x_1, z_1) \leq d_X(x_0, y_1) + d_X(y_1, x_1) + d_X(x_1, z_1) = d_X(x_0, z_1)$.

**Case 2:** The second case to consider is when $1 < i < c$.

1. First assume that $z_i \in X_0$. Then $z_i$ will be at most $i$ in the ordering of $Q_0^{(\text{ball})}$ and hence at most $3i$ in the ordering of $Q_0$. By the induction hypothesis on $X_0$ : $d_T(x_0, z_i) \leq c \log\log(3i) \cdot \text{rad}_{x_0}(X_0) \leq i \cdot \text{rad}_{x_0}(X)$, using that $\text{rad}_{x_0}(X_0) \leq \text{rad}_{x_0}(X)/(8c)$, and that $\log\log(3i) \leq 2i$.
2. Now assume that $\{z_1, \ldots, z_i\} \subseteq X_1$. As $y_1$ is the first in $Q_0$, by the induction hypothesis on $X_0$ and $X_1$ we have that $d_T(x_0, y_1) \leq d_X(x_0, y_1) \leq \text{rad}_{x_0}(X_0)$ and $d_T(x_1, z_i) \leq i \cdot \text{rad}_{x_1}(X_1)$, so

$$\begin{aligned}
d_T(x_0, z_i) &\leq d_T(x_0, y_1) + d_T(y_1, x_1) + d_T(x_1, z_i) \\
&\leq \text{rad}_{x_0}(X_0) + i \cdot \text{rad}_{x_1}(X_1) + d_X(y_1, x_1) \\
&\leq i(\text{rad}_{x_0}(X_0) + d_X(y_1, x_1) + \text{rad}_{x_1}(X_1)) - (i-1)\text{rad}_{x_0}(X_0) \\
&\leq i(1+\epsilon)\text{rad}_{x_0}(X) - (i-1)\text{rad}_{x_0}(X)/(16c) \\
&\leq i \cdot \text{rad}_{x_0}(X) + i \cdot \text{rad}_{x_0}(X)/(170c) - i \cdot \text{rad}_{x_0}(X)/(32c) \\
&\leq i \cdot \text{rad}_{x_0}(X).
\end{aligned}$$

   In the fourth inequality using Claim 2 and that $\text{rad}_{x_0}(X_0) \geq \text{rad}_{x_0}(X)/(16c)$ (note that by the stop condition of hierarchical-star-partition $\text{rad}_{x_0}(X) \geq 16c$, so $\text{rad}_{x_0}(X_0) \geq 1$) and in the fifth that $i - 1 \geq i/2$.

3. Now assume that $z_i \in X_j$ where not all of $z_1, \ldots, z_i$ are in $X_j$ (note that $z_1 \in X_0 \cup X_1$, therefore there is no case for $\{z_1, \ldots, z_i\} \subseteq X_j$ where $j > 1$). First note that $z_i$ must be at most the $i-1$ element in $Q_j$. By the insert sequence to $Q_0^{(\text{reg})}$ we have that $y_j$ is at most the $3i$ element in $Q_0$. Using the induction hypothesis on $X_0$ and $X_j$ we get that

$$\begin{aligned}
d_T(x_0, z_i) &\leq d_T(x_0, y_j) + d_T(y_j, x_j) + d_T(x_j, z_i) \\
&\leq c \log\log(3i) \cdot \text{rad}_{x_0}(X_0) + (i-1) \cdot \text{rad}_{x_j}(X_j) + d_X(y_j, x_j) \\
&\leq (i-1)(\text{rad}_{x_0}(X_0) + d_X(y_j, x_j) + \text{rad}_{x_j}(X_j)) + 5c \cdot \text{rad}_{x_0}(X_0) \\
&\leq (i-1)(1+\epsilon)\text{rad}_{x_0}(X) + 5c \cdot \text{rad}_{x_0}(X)/(8c) \\
&\leq i \cdot \text{rad}_{x_0}(X) - \text{rad}_{x_0}(X) + (i-1) \cdot \text{rad}_{x_0}(X)/(170c) + 5\text{rad}_{x_0}(X)/8 \\
&\leq i \cdot \text{rad}_{x_0}(X).
\end{aligned}$$

   The third inequality follows since $\log\log(3i) \leq \log\log(3c) \leq 5$. The fourth using Claim 2 and that $\text{rad}_{x_0}(X_0) \leq \text{rad}_{x_0}(X)/(8c)$.

**Case 3:** In the third case $i \geq c$.

1. First assume that $z_i \in X_0$. Then $z_i$ will be at most $i$ in the ordering of $Q_0^{(\text{ball})}$, hence at most $3i$ in the ordering of $Q_0$. By the induction hypothesis on $X_0$ we get that

$$d_T(x_0, z_i) \leq c \log\log(3i) \cdot \text{rad}_{x_0}(X_0) \leq 2c \log\log i \cdot \text{rad}_{x_0}(X_0) \leq c \log\log i \cdot \text{rad}_{x_0}(X) \; .$$

   using that for $i \geq c$, $3i < i^2$.

2. Next assume that $z_i \in X_j$ such that $|X_j \cap \{z_1, \ldots, z_i\}| \leq \sqrt{i}$, then $z_i$ will be at most the $\sqrt{i}$ in $Q_j$, and $y_j$ will be at most the $i$-th in $Q_0^{(\text{reg})}$ and hence at most $3i$ in the ordering of $Q_0$. By the induction hypothesis on $X_0$ and $X_j$:

$$\begin{aligned} d_T(x_0, z_i) &\leq d_T(x_0, y_j) + d_T(y_j, x_j) + d_T(x_j, z_i) \\ &\leq c \log\log(3i) \cdot \text{rad}_{x_0}(X_0) + c \log\log(\sqrt{i}) \cdot \text{rad}_{x_j}(X_j) + d_X(y_j, x_j) \\ &\leq c(\log\log i + 1) \cdot \text{rad}_{x_0}(X_0) + c(\log\log i - 1) \cdot \text{rad}_{x_j}(X_j) + d_X(y_j, x_j) \\ &\leq c(\log\log i - 1) \left(\text{rad}_{x_0}(X_0) + d_X(y_j, x_j) + \text{rad}_{x_j}(X_j)\right) + 2c \cdot \text{rad}_{x_0}(X_0) \\ &\leq c(\log\log i - 1)(1+\epsilon)\text{rad}_{x_0}(X) + \text{rad}_{x_0}(X)/4 \\ &\leq c \log\log i \cdot \text{rad}_{x_0}(X) + c \log\log i \cdot \text{rad}_{x_0}(X)/(170c \log\log i) - c \cdot \text{rad}_{x_0}(X) + \text{rad}_{x_0}(X)/4 \\ &\leq c \log\log i \cdot \text{rad}_{x_0}(X), \end{aligned}$$

   the fifth inequality using Claim 2 and that $\text{rad}_{x_0}(X_0) \leq \text{rad}_{x_0}(X)/(8c)$, the sixth that $\epsilon \leq 1/(170c \log\log i)$.

3. The last subcase is where $z_i \in X_j$ such that $|X_j \cap \{z_1, \ldots, z_i\}| > \sqrt{i}$, then $z_i$ will be at most the $i$ in $Q_j$ and by Claim 5 $y_j$ will be at most the $i^{9/10}$ in $Q_0$. Now by the induction hypothesis, for $t \geq 2$

$$\begin{aligned} d_T(x_0, z_i) &\leq d_T(x_0, y_j) + d_X(y_j, x_j) + d_T(x_j, z_i) \\ &\leq c \log\log i^{9/10} \cdot \text{rad}_{x_0}(X_0) + c \log\log i \cdot \text{rad}_{x_j}(X_j) + d_X(y_j, x_j) \\ &\leq c \log\log i(\text{rad}_{x_0}(X_0) + d_X(y_j, x_j) + \text{rad}_{x_j}(X_j)) + c \log(9/10) \cdot \text{rad}_{x_0}(X_0) \\ &\leq c \log\log i \cdot \text{rad}_{x_0}(X) + \epsilon \cdot c \log\log i \cdot \text{rad}_{x_0}(X) - c \cdot \text{rad}_{x_0}(X_0)/10 \\ &\leq c \log\log i \cdot \text{rad}_{x_0}(X) + \text{rad}_{x_0}(X)/170 - \text{rad}_{x_0}(X)/160 \\ &\leq c \log\log i \cdot \text{rad}_{x_0}(X), \end{aligned}$$

   the fourth inequality using Claim 2 and the fifth that $\text{rad}_{x_0}(X_0) \geq \text{rad}_{x_0}(X)/(16c)$ and $\epsilon \leq 1/(170c \log\log i)$.

□

The following claim shows that a portal $y_j$ leading to a point $z_i$ that belongs to a "fat" cone will be located in an improved position in the queue of the central ball $Q_0$.

**Claim 5.** *For any $i \geq 2^{16}$, if $z_i \in X_j$ such that $|X_j \cap \{z_1, \ldots, z_i\}| > \sqrt{i}$ then $y_j$ will be at position at most $i^{9/10}$ in $Q_0$.*

*Proof.* We will show that $y_j$ will be in the first $(3/2)i^{2/3} + 1$ elements of $Q_0^{(\text{fat})}$. Since $i \geq 2^{16}$ it follows that $y_j$ will be in the first $3 \cdot ((3/2)i^{2/3} + 1) < i^{9/10}$ elements of $Q_0$.

Let $y_{i_1}, \ldots, y_{i_s}$ with $i_1 < i_2 < \cdots < i_s$ be a set of $s$ points that were inserted into $Q_0^{(\text{fat})}$ before considering the point $z_i$, we need to show that $s \leq (3/2)i^{2/3}$. Let $z_{i'_1}, \ldots, z_{i'_s}$ be the set of points in $Q$ such that $y_{i_k}$ was inserted because $z_{i'_k} \in X_{i_k}$ and $X_{i_k}$ was a "fat" cone, *i.e.* $|X_{i_k} \cap \{z_1, \ldots, z_{i'_k}\}| \geq \sqrt{i'_k}$. Let $A_{i_k} = X_{i_k} \cap \{z_1, \ldots, z_{i'_k}\}$ denote the set that caused $y_{i_k}$ to enter $Q_0^{(\text{fat})}$, and note that $|A_{i_k}| \geq \sqrt{i'_k} \geq \sqrt{k}$. For any $1 \leq k < \ell \leq s$ we have that $A_{i_k} \cap A_{i_\ell} = \emptyset$, since we do not insert a point $y_{i_\ell}$ that already appear in $Q_0^{(\text{fat})}$, which implies $X_{i_k} \cap X_{i_\ell} = \emptyset$. Note that all the sets $A_{i_k}$ contain points from $z_1,\ldots,z_i$, so we have that $\sum_{k=1}^{s}|A_{i_k}|\le i$. Hence $\sum_{k=1}^{s}\sqrt{k}\le\sum_{k=1}^{s}|A_{i_k}|\le i$. We also bound the sum from below

$$\sum_{k=1}^{s}\sqrt{k}\ge\int_{1}^{s}\sqrt{x}dx=[(2/3)x^{3/2}]_1^s\ge(2/3)s^{3/2},$$

therefore $i\ge(2/3)s^{3/2}$ or $s\le(3/2)i^{2/3}$. ◻

**Corollary 6.** *For any weighted graph $G=(V,E)$ denote by $|V|=n$ and $|E|=m$, invoking* `hierarchical-star-partition` *algorithm on $G$ where in* `star partition` *algorithm we use the* `ImpConeDecompose`$(G,BS(x_0,r_0\cdot\mathrm{rad}(X)),\mathrm{rad}(X)/\log\log n,\log\log n,m)$ *of [10], then we get a single spanning tree $T$ such that*

$$\frac{1}{m}\sum_{(u,v)\in E}\frac{d_T(u,v)}{d_G(u,v)}\le O(\log n\cdot(\log\log n)^3).$$

*The running time is $O(m\log n)$ if $G$ is unweighted and $O(m\log n+n\log^2 n)$ if $G$ is weighted.*

*Proof.* Since our algorithm works in a similar manner to the [10] algorithm, we can use their partitioning method `ImpConeDecompose`, which has a a running time of $O(m)$ if $G$ is unweighted and $O(m+n\log n)$ if $G$ is weighted. The only difference is that in the first iteration ($j=1$), instead of picking an arbitrary portal $x_1$ we pick the node $x_1$ that is first on a shortest path from $x_0$ to the first in the queue $Q$. The average stretch of their cone cutting method is roughly $O(\log n\cdot\log\log n\cdot 1/\epsilon)$ (recall that $\epsilon=1/\log\log n$), and since the radius of our spanning tree increases by $O(\log\log n)$, the corollary follows. It remains to see that our running time is no worse than [10], and indeed it is easy to see that adding the queues increase the run time only by a constant factor. ◻

# 3 Strong Diameter Probabilistic Partitions

$(x,y,r)=$ `cone cut`$(X,x_0,X_0,Y,\epsilon)$:

- Let $p\in Y$ be the point minimizing $\frac{|X|}{|B_{(Y,d_Y)}(z,\epsilon\cdot\mathrm{rad}_{x_0}(X)/16)|}$ over all $z\in Y$; Let $\chi$ denote that minimum;
- Let $(y,x)$ be an edge such that $x\in Y$, $y\in X_0$ and $d_X(x_0,y)+d_X(y,x)+d_Y(x,p)=d_X(x_0,p)$ (*i.e.* $y$ and $x$ lie on some shortest path between $x_0$ and $p$);
- Choose $r\in[\epsilon/4,\epsilon/2]$ according to the following random process:
  - Divide the interval $[\epsilon/4,\epsilon/2]$ into $N=\lceil 2\log\chi\rceil$ equal length intervals $S_1,\ldots,S_N$; Let $h=1$;
  - LOOP: Toss a fair coin; If it turns out head and $h<N$ then let $h=h+1$ and goto LOOP;
  - Choose $r$ uniformly at random from the interval $S_h$.
- Return $(x,y,r)$.

Figure 3: `cone-cut` algorithm

Consider a graph $G=(V,E)$, a connected cluster $X\subseteq V$, $x_0\in X$ and let $\Delta=\mathrm{rad}_{x_0}(X)$. Fix some edge $(u,v)\in E$. Let $X^{(i)}=X^{(i)}(u)$ be a random variable that indicates which cluster contains $u$ in the $i$-th step of the hierarchical application of the star-partition algorithm[4]. In a similar manner let $x_0^{(i)}$ be the random variable indicating the center of the cluster $X^{(i)}$, and when $X^{(i)}$ is partitioned denote the central ball as $X_0^{(i)}$ and cones as $X_1^{(i)},\ldots X_m^{(i)}$ where $m$ is a random variable depending on $X^{(i)}$. Let $\mathcal{E}_j(X^{(i)},u,v)$ be the event that $u,v\in X^{(i)}$ and in the star-partition of the cluster $X^{(i)}$ with center $x_0^{(i)}$ into $X_0^{(i)},\ldots,X_m^{(i)}$, $u\in X_j^{(i)}$, $v\notin X_j^{(i)}$. Let $\mathcal{E}(X^{(i)},u,v)$ be the event that $\exists\, 0\le j\le m$ such that $\mathcal{E}_j(X^{(i)},u,v)$. Some notation:

[4]We abuse notation and think of $X^{(i)}$ as a function to subsets of $X$ (instead of $\mathbb{R}$). We also refer to $X^{(i)}$ as an event.

$\mathbb{E}_{X^{(i)}}[f(X^{(i)})]$ will stand for $\sum_{X'} \Pr[X^{(i)} = X'] f(X')$.

Let $\mathcal{T}$ be the support of the distribution over spanning trees induced by the hierarchical star partition algorithm. Let $\mathcal{T}^{(i)} \subseteq \mathcal{T}$ be the set of spanning trees for which event $\mathcal{E}(X^{(i)}, u, v)$ occurs.

$$\begin{aligned}
\mathbb{E}[d_T(u,v)] &\leq \sum_{i\geq 1} \sum_{T \in \mathcal{T}^{(i)}} \Pr[T] \cdot d_T(u,v) \\
&\leq \sum_{i \geq 1} \mathbb{E}_{X^{(i)}} \left[ \Pr[\mathcal{E}(X^{(i)}, u, v)] \max_{T \in \mathcal{T}^{(i)}} \{d_T(u,v)\} \right] \\
&\leq O(\log\log n) \sum_{i\geq 1} \mathbb{E}_{X^{(i)}} \left[ \Pr[\mathcal{E}(X^{(i)}, u, v)] \cdot \mathrm{rad}_{x_0^{(i)}}(X^{(i)}) \right] .
\end{aligned}$$

The last inequality holds since for any $T \in \mathcal{T}^{(i)}$, $d_T(u,v) \leq d_T(u, x_0^{(i)}) + d_T(x_0, v) \leq 2\mathrm{rad}_{x_0^{(i)}}(T)$ and using Lemma 4 we get that $\mathrm{rad}_{x_0^{(i)}}(T) \leq O(\log\log n \cdot \mathrm{rad}_{x_0^{(i)}}(X^{(i)}))$.

In what follows we bound $\mathbb{E}_{X^{(i)}} \left[ \Pr[\mathcal{E}(X^{(i)}, u, v)] \cdot \mathrm{rad}_{x_0^{(i)}}(X^{(i)}) \right]$. Let $\epsilon = \frac{1}{170c \cdot \log\log|X|}$ and $k = 20c(\ln(1/\epsilon) + 5)$. The main lemma to prove is the following

**Lemma 7.** *For any graph $G = (V, E)$, any edge $(u,v) \in E$, any connected cluster $X^{(i)} \subseteq V$ we have that*

$$\mathbb{E}_{X^{(i)}} \left[ \Pr[\mathcal{E}(X^{(i)}, u, v)] \cdot \mathrm{rad}_{x_0^{(i)}}(X^{(i)}) \right] \leq C \cdot d(u,v)/\epsilon \cdot \left( \mathbb{E}_{X^{(i)}}[\log|X^{(i)}|] - \mathbb{E}_{X^{(i+k)}}[\log|X^{(i+k)}|] \right) .$$

*where $C$ is a universal constant.*

Once this lemma is proved, a telescopic sum argument yields that

$$\begin{aligned}
\mathbb{E}[d_T(u,v)] &\leq O(\log\log n) \sum_{i\geq 1} \mathbb{E}_{X^{(i)}} \left[ \Pr[\mathcal{E}(X^{(i)}, u, v)] \cdot \mathrm{rad}_{x_0}(X^{(i)}) \right] \\
&\leq O(\log\log n) \cdot d(u,v)/\epsilon \sum_{i=1}^{k} \mathbb{E}_{X^{(i)}}[\log|X^{(i)}|] \\
&\leq O(\log n \cdot \log\log n) \cdot d(u,v) \cdot \log(1/\epsilon)/\epsilon \\
&= O(\log n \cdot (\log\log n)^2 \cdot \log\log\log n) \cdot d(u,v) .
\end{aligned}$$

As we stated in the introduction, the algorithm of Figure 3 and proof of Lemma 7 are based on the truncated exponential distribution approach of [5, 1]. The main technical difficulty arises since the space *changes* after each cluster is cut. Dealing with the randomly changing graph raises some additional subtleties in the proof.

We begin with some definitions and an informal description of the algorithm and the proof idea. Fix the edge $(u,v) \in E$, a scale $i$ and $X = X^{(i)}$. Let $Y \subseteq X$ be a random variable indicating that there exists $0 < j \leq m$ such that $Y = Y_{j-1}$ in the star partition of $X$. Define the local growth rate around $x \in Y$ with respect to $Y$ as

$$\chi(X, Y, x) = \frac{|X|}{|B_{Y,d_Y}(x, \epsilon\Delta/16)|}$$

The algorithm for the partition is as follows: Choose a radius for the central ball around $x_0$ from a uniform distribution in a range of size $\approx \Delta/c$. The center $x_1$ is chosen on a shortest path to $z_1$, the first point in the queue, and then the radius for the cone is again sampled from a uniform distribution in a range of size $\approx \epsilon\Delta$. For $j > 1$ the $j$th center $x_j$ is chosen on a shortest path to the point $p_j \in Y_{j-1}$ minimizing $\chi_j = \chi(X, Y_{j-1}, p_j)$, and then the radius of the cone is chosen from a truncated exponential distribution, with parameter $\chi_j$.

Denote the event that $Y = Y_{j-1}$ and $u \in X_j$ as $\mathcal{Z}_j(X, Y, u)$, and let $\mathcal{Z}(X, Y, u)$ be the event that $\exists\, 0 \leq j < m$ such that $\mathcal{Z}_j(X, Y, u)$. Note that fixing $Y_{j-1}$ determines deterministically $p_j$ and therefore also $x_j$ and $\chi_j$. Similarly let $\mathcal{Z}_j(X, Y)$ be the event that $Y = Y_{j-1}$ and $\mathcal{Z}(X, Y)$ the event that $\exists\, 0 \leq j < m$ such that $\mathcal{Z}_j(X, Y)$. Let $N(j)$ be the random variable that is the number of partitions $S_1, \ldots, S_{N(j)}$ of the interval $[\epsilon/4, \epsilon/2]$ for the $j$th cone. Let $0 \leq h(j) \leq N(j)$ be the random variable that is the index of the interval $S_{h(j)}$ from which the radius $r_j$ is uniformly chosen for $X_j$. Some more notation:

$\mathbb{E}_{Y\subseteq X}[f(Y)]$ will stand for $\sum_{Y\subseteq X}\Pr[\mathcal{Z}(X,Y)]\cdot f(Y)$ (we write $\mathbb{E}_Y$ when $X$ is implicit).

$\mathbb{E}_{Y\subseteq X,j}[f(Y)]$ will stand for $\sum_{Y\subseteq X}\Pr[\mathcal{Z}_j(X,Y)]\cdot f(Y)$ (we write $\mathbb{E}_{Y,j}$ when $X$ is implicit).

$\mathbb{E}_{Y\subseteq X,u}[f(Y)]$ will stand for $\sum_{Y\subseteq X}\Pr[\mathcal{Z}(X,Y,u)]\cdot f(Y)$ (we write $\mathbb{E}_{Y,u}$ when $X$ is implicit).

We divide the event $\mathcal{E}(X,u,v)$ into three cases (by symmetry we can define all these events with respect to $u$).

- The first is the event that $u$ falls into one of the first two clusters (the central ball $X_0$ or the first cone $X_1$). This event is denoted by $\mathcal{G}(X,u)$.
- The second is the event that $u$ is contained in cluster $X_j$ for some $j>1$, such that the cone distance between $u$ and the center $x_j$ is in the last interval *i.e.* that $\rho(x_j,u)/\Delta\in S_{N(j)}$. This event is denoted by $\mathcal{F}(X,u)$. We partition the event $\mathcal{F}(X,u)$ using the different values of $j$: For any $j>1$ let $\mathcal{F}_j(X,u)$ be the event that $\rho(x_j,u)/\Delta\in S_{N(j)}$, and note that $\mathcal{F}(X,u)$ is simply that there exists $j>1$ such that $\mathcal{F}_j(X,u)$ and also $u\in X_j$.
- The third is the completion of the first two events, that the cluster $X_j$ containing $u$ has $j>1$ and $\rho(x_j,u)/\Delta\notin S_{N(j)}$.

The probability of the first event can be bounded simply by the inverse of the range from which the radius is drawn, so we obtain probability at most $\approx\frac{d(u,v)}{\epsilon\Delta}$.

For the second event we note that reaching the tail of the exponential distribution requires that $N-1$ fair coin tosses turned out head, which is bounded by $\approx\frac{1}{2^N}\approx\frac{1}{\chi_j^2}$, then since we choose uniformly from the last interval, the probability that we separate $u,v$ is $\approx\frac{\log\chi_j\cdot d(u,v)}{\epsilon\Delta\chi_j^2}\le\frac{d(u,v)}{\epsilon\Delta\chi_j}$. Since the parameter $\chi_j$ is a random variable which depends on the previous cone cuts, the proof becomes a bit more involved as we need to give a different bound for every possible $Y=Y_{j-1}$. We show that for every star-partition $\sum_{j>1}\chi_j^{-1}\le 1$, hence this also holds in expectation and the second event probability is bounded by $\approx\frac{d(u,v)}{\epsilon\Delta}$. This is shown in Claim 8

Bounding the third event relies on the memoryless property of the exponential distribution. The major technical difficulty is that the bound we show depends on the parameter $\chi$. Hence we can only show the bound given some subspace $Y$ from which we cut the next cone. The bound on the probability obtained here is $\approx\frac{\log\chi\cdot d(u,v)}{\epsilon\Delta}$. This is shown in Claim 9.

The last step is to sum over all scales $i$, and use a telescopic sum argument on the expectation of the values of the $\log\chi$ showing that they sum to $O(\log(1/\epsilon)\cdot\log n)$. This is shown in the proof of Lemma 7.

**Claim 8.** *For any cluster $X\subseteq V$, edge $u,v\in X$, $(u,v)\in E$, we have*

$$\Pr[\mathcal{F}(X,u)\wedge\mathcal{E}(X,u,v)]\le 48d(u,v)/(\epsilon\Delta)\,.$$

*Proof.* Note that we can only bound the probability of event such as $\mathcal{E}_j(X,u,v)$ given that some $Y=Y_{j-1}$ is fixed *i.e.* that event $\mathcal{Z}_j(X,Y)$ occurred (because the parameters $x_j$ and $\chi_j$ that govern the next cone creation are random variables depending on $Y$. So fix some $Y=Y_{j-1}$ and note that indeed $p_j$, $x_j$ and $\chi_j=\chi(X,Y,p_j)$ are determined deterministically.

$$\begin{aligned}
&\Pr[\mathcal{F}(X,u)\wedge\mathcal{E}(X,u,v)]\\
&\quad=\Pr[\exists j>1,\mathcal{F}_j(X,u)\wedge\mathcal{E}_j(X,u,v)]\\
&\quad\le\sum_{j\ge2}\Pr[\mathcal{E}_j(X,u,v)\mid\mathcal{F}_j(X,u)]\\
&\quad=\sum_{j\ge2}\sum_{Y\subseteq X}\Pr[Z_j(X,Y)]\cdot\Pr[\mathcal{E}_j(X,u,v)\mid\mathcal{F}_j(X,u)\wedge\mathcal{Z}_j(X,Y)]\\
&\quad=\sum_{j\ge2}\mathbb{E}_{Y,j}\left[\Pr[\mathcal{E}_j(X,u,v)\mid\mathcal{F}_j(X,u)]\right]
\end{aligned}$$

The first equation holds since the probability to be cut by a cluster whose radius is "large" is the probability that some cluster $X_j$ with large radius separates $u, v$. The first inequality holds by the union bound and the second equation since for every event $A$ and pairwise disjoint events $B_1, \ldots, B_\ell$ with $\sum_{i=1}^{\ell} \Pr[B_i] = 1$ it holds that $\Pr[A] = \sum_{i=1}^{\ell} \Pr[B_i] \cdot \Pr[A \mid B_i]$. Here the events $B$ are $\mathcal{Z}_j(X, Y)$ which are disjoint for different subgraphs $Y$. Note that events $\mathcal{F}_j(X, u)$ and $\mathcal{Z}_j(X, Y)$ tell us nothing of the radius of the next cone $X_j$, therefore the probability of $\mathcal{E}_j(X, u, v)$ given the subspace $Y_{j-1}$ and that $\rho(x_j, u)/\Delta \in S_{N(j)}$ (where $\rho = \rho(X, Y \cup X_0, d', x_0, x_j)$ is the cone metric), is the probability that $h(j) = N(j)$ (recall that the random variable $h(j)$ is the index of the interval $S_{h(j)}$ from which the radius is uniformly chosen for $X_j$) and that the uniform choice in the interval $S_{N(j)}$ hits the place that separates $u, v$. To bound the first one

$$\Pr[h(j) = N(j)] = 2^{-(N(j)-1)} \leq 2^{-2\log \chi_j + 2} = 4/\chi^2,$$

and the probability of the second event is $\frac{d(u,v)}{\Delta |S_{N(j)}|}$. Note that $|S_{N(j)}| = \frac{\epsilon}{4\lceil 2\log \chi_j \rceil} \geq \frac{\epsilon}{8\log \chi_j + 4} \geq \min\{1, \frac{1}{\log \chi_j}\} \cdot \frac{\epsilon}{12}$. These two events are independent, hence

$$\begin{aligned}
\Pr[\mathcal{F}(X, u) \wedge \mathcal{E}(X, u, v)] &\leq \frac{48 d(u, v)}{\epsilon \cdot \Delta} \sum_{j \geq 1} \mathbb{E}_{Y,j}\left[\max\left\{\frac{1}{\chi_j^2}, \frac{\log \chi_j}{\chi_j^2}\right\}\right] \\
&\leq \frac{48 d(u, v)}{\epsilon \cdot \Delta} \sum_{j \geq 1} \mathbb{E}_{Y,j}[\chi_j^{-1}]
\end{aligned}$$

For any $\bar{Y} = (\bar{Y}_1, \bar{Y}_2, \ldots, \bar{Y}_n) \subset X^n$ let $\mathcal{Z}(\bar{Y})$ be the event $\bigwedge_{1 \leq j \leq n} \mathcal{Z}(X, \bar{Y}_j, j)$ (where $\bar{Y}_j$ is the $j$th component of $\bar{Y}$). Observe that for any $j$ and $Y \subset X$ we have $\Pr[\mathcal{Z}(X, Y, j)] = \sum_{\bar{Y} \subset X^n, \bar{Y}_j = Y} \Pr[\mathcal{Z}(\bar{Y})]$. Therefore

$$\begin{aligned}
\sum_{j > 1} \mathbb{E}_{Y,j}[\chi_j^{-1}] &= \sum_{j > 1} \sum_{Y \subseteq X} \Pr[Z(X, Y, j)] \cdot \chi_j^{-1} \\
&= \sum_{j \geq 1} \sum_{\bar{Y} \subset X^n} \Pr[\mathcal{Z}(\bar{Y})] \cdot \chi_j^{-1} \\
&= \sum_{\bar{Y} \subset X^n} \Pr[\mathcal{Z}(\bar{Y})] \sum_{j \geq 1} \chi_j^{-1}
\end{aligned}$$

Now it is enough to show that for any $X_0, X_1, \ldots, X_m$ that may occur in the start-partition algorithm (*i.e.* $\Pr[\mathcal{Z}(\bar{Y})] > 0$, given that $\bar{Y}_j = X \setminus \bigcup_{\ell < j} X_\ell$) we have $\sum_{j=1}^{m} \chi_j^{-1} \leq 1$. This holds because for any $2 \leq \ell < j \leq m$ we have that $B_{Y_\ell, d_{Y_\ell}}(p_\ell, \epsilon\Delta/16) \subseteq X_\ell$, and $Y_j \cap X_\ell = \emptyset$, *i.e.* $B_{Y_\ell, d_{Y_\ell}}(p_i, \epsilon\Delta/16) \cap B_{Y_j, d_{Y_j}}(p_j, \epsilon\Delta/16) = \emptyset$. Therefore

$$\sum_{j=1}^{m} \chi_j^{-1} \leq |X|^{-1} \sum_{j=1}^{m} B_{Y_j, d_j}(p_j, \epsilon\Delta/16) \leq 1.$$

□

**Claim 9.** *For any cluster $X \subseteq V$, edge $u, v \in X$, $(u, v) \in E$, subgraph $Y \subset X$ we have*

$$\Pr[\mathcal{E}(X, u, v) \wedge \neg\mathcal{F}(X, u) \mid \neg\mathcal{G}(X, u) \wedge \mathcal{Z}(X, Y, u)] \leq 12 d(u, v) \max\{1, \log \chi(X, Y, u)\}/(\epsilon \cdot \Delta)$$

*Proof.* If $d(u, v) \geq \epsilon \cdot \Delta/12$ the the claim is trivial, so assume it is smaller. Let $j > 1$ be such that the next cone to be cut is $X_j$ (the value of $j$ is not relevant, we fix it in order to simplify the notation), and recall that fixing $Y = Y_{j-1}$ determines deterministically $p_j$, $x_j$ and $\chi_j$. Let $\rho = \rho(X_0 \cup Y, Y, x_0, x_j)$ be the appropriate cone metric on $Y$ by which the next cone is cut.

$$\begin{aligned}
\Pr[\mathcal{E}(X, u, v) \wedge \neg\mathcal{F}(X, u) \mid \mathcal{Z}(X, Y, u)] &\leq \Pr[\mathcal{E}_j(X, u, v) \wedge \neg\mathcal{F}_j(X, u) \mid \mathcal{Z}(X, Y, u) \wedge \mathcal{Z}(X, Y)] \\
&\leq \Pr[\mathcal{E}_j(X, u, v) \mid \rho(x_j, u)/\Delta \notin S_{N(j)} \wedge \mathcal{Z}(X, Y, u) \wedge \mathcal{Z}(X, Y)] \\
&\leq \frac{\Pr[\mathcal{E}_j(X, u, v) \mid \rho(x_j, u)/\Delta \notin S_{N(j)} \wedge \mathcal{Z}(X, Y)]}{\Pr[\mathcal{Z}(X, Y, u) \mid \rho(x_j, u)/\Delta \notin S_{N(j)} \wedge \mathcal{Z}(X, Y)]}
\end{aligned}$$

The first inequality holds since event $\mathcal{Z}(X,Y,u)$ implies that $u \in X_j$ so the events $\mathcal{E}(X,u,v)$ and $\mathcal{E}_j(X,u,v)$ are equivalent (the same holds for $\neg\mathcal{F}(X,u)$), and because $\mathcal{Z}(X,Y,u) \subseteq \mathcal{Z}(X,Y)$. The second is by the definition of $\mathcal{F}(X,u)$ (given that $u \in X_j$ it cannot be that $\rho(x_j,u)/\Delta$ falls in the interval $S_{N(j)}$), and since for any events $A, B$, $\Pr[A \wedge B] \leq \Pr[A \mid B]$. The third is by Bayes rule and since $\mathcal{E}_j(X,u,v) \wedge \mathcal{Z}(X,Y,u) = \mathcal{E}_j(X,u,v)$. Let $\ell$ be such that $\rho(x_j,u)/\Delta \in S_\ell$.

First we bound the denominator, noting that there is no prior information given about the distribution for the next choice of radius. Since $\ell < N(j)$ we can bound $\Pr[\mathcal{Z}(X,Y,u) \mid \rho(x_j,u)/\Delta \notin S_{N(j)} \wedge \mathcal{Z}(X,Y)] \geq 2^{-\ell}$, since with this probability the radius for the cone $X_j$ will be chosen from $S_m \cdot \Delta$ with $m > \ell$ so it will large enough to contain $u$. The numerator $\Pr[\mathcal{E}_j(X,u,v) \mid \rho(x_j,u)/\Delta \notin S_{N(j)} \wedge \mathcal{Z}(X,Y)]$ can be bounded by $\frac{1}{2^{\ell-1}} \cdot \frac{1}{2} \cdot \frac{d(u,v)}{\Delta|S_\ell|}$, which is the probability that we reach the $\ell$-th interval, not continue to the next one (note that the next interval exists because $\ell < N(j)$) and when choosing $r_j$ uniformly from $S_\ell$, it happens to be the place that separates $u, v$. The probability for the first event is $2^{-(\ell-1)}$, the second is $1/2$, and the third is $\frac{d(u,v)}{\Delta|S_\ell|}$. Since $|S_\ell| \geq \min\{1, \frac{1}{\log \chi_j}\} \cdot \frac{\epsilon}{12}$ it follows that $\Pr[\mathcal{E}_j(X,u,v) \mid \rho(x_j,u)/\Delta \notin S_{N(j)} \wedge \mathcal{Z}(X,Y)] \leq \frac{12d(u,v)\max\{1,\log\chi_j\}}{\epsilon\cdot\Delta\cdot 2^\ell}$. We conclude that

$$\Pr[\mathcal{E}(X,u,v) \wedge \neg\mathcal{F}(X,u) \mid \mathcal{Z}(X,Y,u)] \leq \frac{12d(u,v)\max\{1,\log\chi_j\}}{\epsilon\cdot\Delta}.$$

□

*Proof of Lemma 7.* Fix any $i \geq 1$ and $X^{(i)} = X^{(i)}(u)$. As described before we partition the event $\mathcal{E}(X^{(i)},u,v)$, given a fixed cluster $X^{(i)}$ into the three cases.

$$\begin{aligned}
&\Pr[\mathcal{E}(X^{(i)},u,v)] \\
&= \Pr[\mathcal{E}(X^{(i)},u,v) \wedge \mathcal{F}(X^{(i)},u)] + \Pr[\mathcal{E}(X^{(i)},u,v) \wedge \neg\mathcal{F}(X^{(i)},u)] \\
&= \Pr[\mathcal{E}(X^{(i)},u,v) \wedge \mathcal{F}(X^{(i)},u)] + \Pr[\mathcal{E}(X^{(i)},u,v) \wedge \mathcal{G}(X^{(i)},u)] + \Pr[\mathcal{E}(X^{(i)},u,v) \wedge \neg\mathcal{F}(X^{(i)},u) \wedge \neg\mathcal{G}(X^{(i)},u)]
\end{aligned}$$

The last equality holds since event $\mathcal{G}(X^{(i)},u)$ implies that $\neg\mathcal{F}(X^{(i)},u)$. We claim that the following hold:

$$\Pr[\mathcal{E}(X^{(i)},u,v) \wedge \mathcal{F}(X^{(i)},u) \mid X^{(i)}] \leq 48d(u,v)/(\epsilon\Delta) \tag{2}$$
$$\Pr[\mathcal{E}(X^{(i)},u,v) \wedge \mathcal{G}(X^{(i)},u) \mid X^{(i)}] \leq 5d(u,v)/(\epsilon\Delta) \tag{3}$$
$$\Pr[\mathcal{E}(X^{(i)},u,v) \wedge \neg\mathcal{F}(X^{(i)},u) \wedge \neg\mathcal{G}(X^{(i)},u) \mid X^{(i)}] \leq 12d(u,v)/(\epsilon\Delta) \cdot \mathbb{E}_{Y,u}[\max\{1, \log\chi(X,Y,u)\}] \tag{4}$$

(2) holds directly from Claim 8. (3) since the radius of the central ball is chosen uniformly from interval of length $\Delta/(16c) \geq \epsilon\Delta$, and for the first cone from interval of length $\epsilon\Delta/4$. (4) holds by using Claim 9 and writing

$$\begin{aligned}
\Pr[\mathcal{E}(X^{(i)},u,v) \wedge \neg\mathcal{F}(X^{(i)},u) \wedge \neg\mathcal{G}(X^{(i)},u)] &\leq \mathbb{E}_{Y,u}\left[\Pr[\mathcal{E}(X^{(i)},u,v) \wedge \neg\mathcal{F}(X^{(i)},u) \mid \neg\mathcal{G}(X^{(i)},u)]\right] \\
&\leq \frac{12d(u,v)}{\epsilon\cdot\Delta}\mathbb{E}_{Y,u}[\max\{1, \log\chi(X,Y,u)\}]
\end{aligned}$$

Combining these three equation yields that for $C = 65$

$$\Pr[\mathcal{E}(X^{(i)},u,v)] \leq C \cdot d(u,v)/(\epsilon\Delta) \cdot \mathbb{E}_{Y,u}[\max\{1, \log\chi(X,Y,u)\}] .$$

Recall that $k = 20c(\ln(1/\epsilon)+5)$, and Corollary 3 suggests that for any cluster $X$ and any $j \geq 0$ that $\mathrm{rad}_{x_j}(X_j) \leq (1-1/(20c))\mathrm{rad}_{x_0}(X)$, hence for any event $X^{(i+k)}$, given that $X^{(i)}$ happened

$$\mathrm{rad}(X^{(i+k)}) \leq (1-1/(20c))^k \cdot \mathrm{rad}(X^{(i)}) \leq \epsilon \cdot \mathrm{rad}(X^{(i)})/32,$$

therefore $\mathrm{diam}(X^{(i+k)}) \leq \epsilon\cdot\mathrm{rad}(X^{(i)})/16$ and by definition $u \in X^{(i+k)}$, so fixing any $Y$ such that event $\mathcal{Z}(X^{(i)},Y,u)$ occurred then if $X^{(i+k)} \subseteq Y$ also $X^{(i+k)} \subseteq B_{Y,d_Y}(u, \epsilon \cdot \mathrm{rad}(X^{(i)})/16)$.

$$\begin{aligned}
\mathbb{E}_{Y,u}[\log \chi(X^{(i)}, Y, u)] &= \log |X^{(i)}| - \mathbb{E}_{Y,u}[\log |B_{Y,d_Y}(u, \epsilon \cdot \mathrm{rad}(X^{(i)})/16|] \\
&\leq \log |X^{(i)}| - \mathbb{E}_{Y,u}\left[\sum_{X^{(i+k)} \subseteq Y} \Pr[X^{(i+k)} \mid \mathcal{Z}(X^{(i)}, Y, u)] \cdot \log |X^{(i+k)}|\right] \\
&= \log |X^{(i)}| - \sum_{X^{(i+k)} \subseteq X^{(i)}} \Pr[X^{(i+k)} \mid X^{(i)}] \cdot \log |X^{(i+k)}|
\end{aligned}$$

We conclude that

$$\begin{aligned}
\mathbb{E}_{X^{(i)}}\left[\Pr[\mathcal{E}(X^{(i)}, u, v)]\right] &\leq \mathbb{E}_{X^{(i)}}\left[\log |X^{(i)}| - \sum_{X^{(i+k)} \subseteq X^{(i)}} \Pr[X^{(i+k)} \mid X^{(i)}] \cdot \log |X^{(i+k)}|\right] \\
&= \mathbb{E}_{X^{(i)}}[\log |X^{(i)}|] - \left[\sum_{X^{(i)}} \Pr[X^{(i)}] \sum_{X^{(i+k)} \subseteq X^{(i)}} \Pr[X^{(i+k)} \mid X^{(i)}] \cdot \log |X^{(i+k)}|\right] \\
&= \mathbb{E}_{X^{(i)}}[\log |X^{(i)}|] - \mathbb{E}_{X^{(i+k)}} \log |X^{(i+k)}|
\end{aligned}$$

□

## Acknowledgments

We would like to thank Michael Elkin for initial discussions on the problem, Harald Räcke and Yuval Emek for comments on a preliminary version.

# A Extending to Weighted Graphs

In order for our algorithm to work for general weighted graphs, we will make the following change: After choosing the points $x_j$, $y_j$ in the cone-cut algorithm, create an imaginary point $y'_j$ which lies on the edge $y_j, x_j$ such that $d(x_0, y'_j) = \mathrm{rad}_{x_0}(X_0)$, then return the point $y'_j$. Note that then the inequality $\mathrm{rad}_{x_0}(X_0) + d(y'_j, x_j) + \mathrm{rad}_{x_j}(X_j) \leq (1+\epsilon)\mathrm{rad}_{x_0}(X)$ will hold, which is the only place we used the unweighted property of $G$. With a slight change to the algorithm the number of imaginary points added is at most the number of edges in the original graph $G$. This is because the point $y'_j$ is connected only to $y_j$ in the central ball $X_0$, so if in the recursion depth when cutting a cluster $\hat{X}$, the edge is cut by the central ball $\hat{X}_0$, then the cone $\hat{X}_\ell$ created will contain only one point - $y_j = x_\ell$, so in such a case it will hold that $\mathrm{rad}_{x_0}(\hat{X}_0) + d(y_j, y'_j) + \mathrm{rad}_{x_\ell}(\hat{X}_\ell) \leq (1+\epsilon)\mathrm{rad}_{x_0}(\hat{X})$, and we will not add another imaginary point.

The other change to the algorithm is contraction of small edges, following [10]. Let $G = (V, E)$ be the original graph of size $n$. At every recursive step of `hierarchical star partition` for a cluster $X$ with $\Delta = \mathrm{rad}(X)$ we contract all edges shorter than $c\Delta/n$ for a constant $c$. Then these small edges will not be cut - it guarantees that every edge is at risk in at most $O(\log n)$ recursive steps. It remains to show that the radius does not increase - note that adding back all these edges will increase the radius by at most $c\Delta$, and also note that our inductive proof actually has a slack of $c'\Delta$ , *i.e.* if we need to bound $d_T(x_0, z_i)$ by $i \cdot \Delta$ then we actually show that $d_T(x_o, z_i) \leq i \cdot \Delta - c'\Delta$. Now choosing $c < c'$ will guarantee that even after expanding back all the edges we contracted the radius bound still holds. The last issue is the choice of portals in the expanded graph. If $\hat{x_j}$ is the super node in the $j$-th portal (recall that $y'_j$ is an added imaginary point), we choose $x_j \in \hat{x_j}$ which is connected to some vertex in $\hat{y_j}$ and also lies on the shortest path from $x_0$ to $z_i$.

# B Improving the stretch slightly

The factor of $c \log \log i$ that was chosen as a bound on the radius increase in Lemma 4 was actually arbitrary. In fact we can replace it with almost any other monotone increasing function of $i$, the position in the queue. In order to optimize (asymptotically) the stretch, we take a very slowly increasing function of $i$, using the following definitions: Recall that $\log^{(0)} n = n$ and for any integer $1 \le t \le \log^* n$, $\log^{(t)} n = \log\left(\log^{(t-1)} n\right)$. We use $\log^* n = \min\{t \mid 1 \le \log^{(t)} n < 2\}$. For any integer $1 \le t \le \log^* n$ let $\varphi_t(n) = \prod_{k=2}^{t} \log^{(k)} n$, (when $t = 1$ let $\varphi_1(n) = 1$).

The following two technical claims are proven in Appendix C.

**Claim 10.** *For any $0 \le a \le 1$, $i \ge 4$ and integer $2 \le t \le \log^* i$,*

$$\log^{(t)}(i^a) \le \log^{(t)} i + (\log a)/\varphi_{t-1}(i).$$

**Claim 11.** *For any $a \ge 1$, $i \ge 16$ and integer $2 \le t \le \log^* i$,*

$$\log^{(t)}(ai) \le \log^{(t)} i + (2\log a)/\log i.$$

The parameter $c$ that was a constant can now be arbitrary number $c \ge 2^{18}$, *i.e.* it can be a function of $|X|$. We also use a different value of $\epsilon = \frac{1}{170c \cdot \varphi_t(n)}$ for the star partition. Now the lemma that gives a tighter bound on the radius is the following:

**Lemma 12.** *Let $1 \le t \le \log^* c$ be an integer. Let $(X, d)$ be the metric derived from an unweighted graph $G = (V, E)$, $x_0 \in X$ and $Q = (z_1, \ldots, z_{|X|-1})$ any ordering of $X \setminus \{x_0\}$, also let $T$ be the spanning tree of $G$ returned by the algorithm* `hierarchical-star-partition`$(X, x_0, Q)$ *with parameter $\epsilon = \epsilon(X, c, t) = \frac{1}{90c\varphi_t(|X|)}$, then*

$$d_T(x_0, z_i) \le \begin{cases} d(x_0, z_i) & i = 1 \\ i \cdot \mathrm{rad}_{x_0}(X) & 1 < i < c \\ c \cdot \log^{(t)} i \cdot \mathrm{rad}_{x_0}(X) & \textit{otherwise} \end{cases}$$

*Proof.* The proof is by induction on the radius of $X$. Note that Corollary 3 guarantees that for all $j = 0, \ldots, m$ we have $1 \le \mathrm{rad}_{x_j}(X_j) < \mathrm{rad}_{x_0}(X)$.

**Case 1:** The case $i = 1$ is identical to Lemma 4.

**Case 2:** The second case to consider is when $1 < i < c$.

1. First assume that $z_i \in X_0$. Then $z_i$ will be at most $i$ in the ordering of $Q_0^{(\mathrm{ball})}$ and hence at most $3i$ in the ordering of $Q_0$. By the induction hypothesis on $X_0$ : $d_T(x_0, z_i) \le c\log^{(t)}(3i) \cdot \mathrm{rad}_{x_0}(X_0) \le i \cdot \mathrm{rad}_{x_0}(X)$, using that $\mathrm{rad}_{x_0}(X_0) \le \mathrm{rad}_{x_0}(X)/(4c)$, and that $\log^{(t)}(3i) \le 2i$.

2. Next assume that $\{z_1, \ldots, z_i\} \subseteq X_1$. As $y_1$ is the first in $Q_0$, by the induction hypothesis on $X_0$ and $X_1$ we have that $d_T(x_0, y_1) \le d_X(x_0, y_1) \le \mathrm{rad}_{x_0}(X_0)$ and $d_T(x_1, z_i) \le i \cdot \mathrm{rad}_{x_1}(X_1)$, so

$$\begin{aligned} d_T(x_0, z_i) &\le d_T(x_0, y_1) + d_T(y_1, x_1) + d_T(x_1, z_i) \\ &\le \mathrm{rad}_{x_0}(X_0) + i \cdot \mathrm{rad}_{x_1}(X_1) + d(y_1, x_1) \\ &\le i(\mathrm{rad}_{x_0}(X_0) + d(y_1, x_1) + \mathrm{rad}_{x_1}(X_1)) - (i-1)\mathrm{rad}_{x_0}(X_0) \\ &\le i(1+\epsilon)\mathrm{rad}_{x_0}(X) - (i-1)\mathrm{rad}_{x_0}(X)/(16c) \\ &\le i \cdot \mathrm{rad}_{x_0}(X) + i \cdot \mathrm{rad}_{x_0}(X)/(170c) - i \cdot \mathrm{rad}_{x_0}(X)/(32c) \\ &\le i \cdot \mathrm{rad}_{x_0}(X). \end{aligned}$$

In the fourth inequality using that $\mathrm{rad}_{x_0}(X_0) \ge \mathrm{rad}_{x_0}(X)/(16c)$, in the fifth that $i/(i-1) \le 2$.

3. Now assume that $z_i \in X_j$ where not all of $z_1, \dots, z_i$ are in $X_j$. This case further subdivides to two main cases, the second one divides to two subcases (this complication arises since $c$ is not a constant anymore).

   (a) If $i \le c/4$: First note that $z_i$ must be at most the $i-1$ element in $Q_j$. By the insert sequence to $Q_0^{(\text{reg})}$ we have that $y_j$ is at most the $3i < c$ element in $Q_0$. Using the induction hypothesis on $X_0$ and $X_j$ we get that

$$\begin{aligned}
d_T(x_0, z_i) &\le d_T(x_0, y_j) + d_T(y_j, x_j) + d_T(x_j, z_i) \\
&\le i \cdot \mathrm{rad}_{x_0}(X_0) + (i-1) \cdot \mathrm{rad}_{x_j}(X_j) + d(y_j, x_j) \\
&\le (i-1)(\mathrm{rad}_{x_0}(X_0) + d(y_j, x_j) + \mathrm{rad}_{x_j}(X_j)) + \mathrm{rad}_{x_0}(X_0) \\
&\le (i-1)(1+\epsilon)\mathrm{rad}_{x_0}(X) + \mathrm{rad}_{x_0}(X)/(8c) \\
&\le i \cdot \mathrm{rad}_{x_0}(X) - \mathrm{rad}_{x_0}(X) + (i-1) \cdot \mathrm{rad}_{x_0}(X)/(170c) + \mathrm{rad}_{x_0}(X)/(8c) \\
&\le i \cdot \mathrm{rad}_{x_0}(X).
\end{aligned}$$

   (b) Otherwise $i > c/4$, then there are two cases:

   - If $|X_j \cap \{z_1, \dots, z_i\}| \le \sqrt{i}$ then $z_i$ will be at most the $\sqrt{i}$ in $Q_j$ and $y_j$ will be at most the $3i$ in $Q_0$. Note that for $c > 100$, $\sqrt{i} < i/2$, and also $\log^{(t)}(3i) < i$ for all $t \ge 1$, hence

$$\begin{aligned}
d_T(x_0, z_i) &\le d_T(x_0, y_j) + d_T(y_j, x_j) + d_T(x_j, z_i) \\
&\le c\log^{(t)}(3i) \cdot \mathrm{rad}_{x_0}(X_0) + (i/2) \cdot \mathrm{rad}_{x_j}(X_j) + d(y_j, x_j) \\
&\le c \cdot i \cdot \mathrm{rad}_{x_0}(X_0) + (i/2) \cdot \mathrm{rad}_{x_0}(X) \\
&\le i \cdot \mathrm{rad}_{x_0}(X)/8 + (i/2)\mathrm{rad}_{x_0}(X) \\
&\le i \cdot \mathrm{rad}_{x_0}(X),
\end{aligned}$$

     using that $\mathrm{rad}_{x_0}(X_0) \le \mathrm{rad}_{x_0}(X)/(8c)$.
   - If $|X_j \cap \{z_1, \dots, z_i\}| > \sqrt{i}$ then $z_i$ will be at most the $i$-th in $Q_j$ and by Claim 5 $y_j$ will be at most the $i^{9/10}$ in $Q_0$. Note that $i^{9/10} < i/2$, then by the induction hypothesis

$$\begin{aligned}
d_T(x_0, z_i) &\le d_T(x_0, y_j) + d_T(y_j, x_j) + d_T(x_j, z_i) \\
&\le (i/2) \cdot \mathrm{rad}_{x_0}(X_0) + i \cdot \mathrm{rad}_{x_j}(X_j) + d(y_j, x_j) \\
&\le i \cdot (\mathrm{rad}_{x_0}(X_0) + d(y_j, x_j) + \cdot\mathrm{rad}_{x_j}(X_j)) - (i/2) \cdot \mathrm{rad}_{x_0}(X_0) \\
&\le i \cdot \mathrm{rad}_{x_0}(X) + \epsilon \cdot i \cdot \mathrm{rad}_{x_0}(X) - i \cdot \mathrm{rad}_{x_0}(X)/(32c) \\
&\le i \cdot \mathrm{rad}_{x_0}(X),
\end{aligned}$$

     using that $\mathrm{rad}_{x_0}(X_0) \ge \mathrm{rad}_{x_0}(X)/(16c)$.

**Case 3:** The third case when $i \ge c$:

1. If $z_i \in X_0$ then it will be at most $i$ in the ordering of $Q_0^{(\text{ball})}$ hence at most $3i$ in the ordering of $Q_0$. By the induction hypothesis on $X_0$ we get that

$$d_T(x_0, z_i) \le c\log^{(t)}(3i) \cdot \mathrm{rad}_{x_0}(X_0) \le 2c\log^{(t)}(i) \cdot \mathrm{rad}_{x_0}(X_0) \le c\log^{(t)} i \cdot \mathrm{rad}_{x_0}(X),$$

   using that for $i \ge c$, $3i < i^2$ hence $\log^{(t)}(3i) \le 2\log^{(t)} i$ for all $t$.

2. The second case is when $z_i \in X_j$ such that $|X_j \cap \{z_1, \dots, z_i\}| \le \sqrt{i}$, then $z_i$ will be at most the $\sqrt{i}$ in $Q_j$, and $y_j$ will be at most the $i$-th in $Q_0^{(\text{reg})}$ and hence at most $3i$ in the ordering of $Q_0$. By the induction hypothesis on

$X_0$ and $X_j$, for $t \geq 2$ :

$$\begin{aligned}
d_T(x_0, z_i) &\leq d_T(x_0, y_j) + d_T(y_j, x_j) + d_T(x_j, z_i) \\
&\leq c \log^{(t)}(3i) \cdot \mathrm{rad}_{x_0}(X_0) + c \log^{(t)}(\sqrt{i}) \cdot \mathrm{rad}_{x_j}(X_j) + d(y_j, x_j) \\
&\leq c(\log^{(t)} i + 4/\log i) \cdot \mathrm{rad}_{x_0}(X_0) + c(\log^{(t)} i - 1/\varphi_{t-1}(i)) \cdot \mathrm{rad}_{x_j}(X_j) + d(y_j, x_j) \\
&= c(\log^{(t)} i - 1/\varphi_{t-1}(i)) \left(\mathrm{rad}_{x_0}(X_0) + d(y_j, x_j) + \mathrm{rad}_{x_j}(X_j)\right) + c(4/\log i + 1/\varphi_{t-1}(i)) \cdot \mathrm{rad}_{x_0}(X_0) \\
&\leq c(\log^{(t)} i - 1/\varphi_{t-1}(i))(1+\epsilon)\mathrm{rad}_{x_0}(X) + (1/(2\log i) + 1/(8\varphi_{t-1}(i)))\mathrm{rad}_{x_0}(X) \\
&\leq c \log^{(t)} i \cdot \mathrm{rad}_{x_0}(X) + \mathrm{rad}_{x_0}(X) \cdot \log^{(t)} i/(170\varphi_t(i)) - \mathrm{rad}_{x_0}(X)/\varphi_{t-1}(i) + 3\mathrm{rad}_{x_0}(X)/(4\varphi_{t-1}(i)) \\
&\leq c \log^{(t)} i \cdot \mathrm{rad}_{x_0}(X) + \mathrm{rad}_{x_0}(X)/(170\varphi_{t-1}(i)) - \mathrm{rad}_{x_0}(X)/(4\varphi_{t-1}(i)) \\
&\leq c \log^{(t)} i \cdot \mathrm{rad}_{x_0}(X),
\end{aligned}$$

the third inequality using Claim 10 and Claim 11. The fifth inequality holds since for every $s \geq 1$, $\log i \geq \varphi_s(i)$, and so $\mathrm{rad}_{x_0}(X)/\log i \leq \mathrm{rad}_{x_0}(X)/\varphi_{t-1}(i)$, and the sixth because $\log^{(t)} i/\varphi_t(i) = 1/\varphi_{t-1}(i)$.

In a similar manner, it can shown that the same holds for $t = 1$.

3. The last case is where $z_i \in X_j$ such that $|X_j \cap \{z_1, \ldots, z_i\}| > \sqrt{i}$, then $z_i$ will be at most the $i$ in $Q_j$ and by Claim 5 $y_j$ will be at most the $i^{9/10}$ in $Q_0$. Now by the induction hypothesis, for $t \geq 2$

$$\begin{aligned}
d_T(x_0, z_i) &\leq d_T(x_0, y_j) + d(y_j, x_j) + d_T(x_j, z_i) \\
&\leq c \log^{(t)} i^{9/10} \cdot \mathrm{rad}_{x_0}(X_0) + c \log^{(t)} i \cdot \mathrm{rad}_{x_j}(X_j) + d(y_j, x_j) \\
&\leq c \log^{(t)} i(\mathrm{rad}_{x_0}(X_0) + d(y_j, x_j) + \mathrm{rad}_{x_j}(X_j)) + c\log(9/10)/\varphi_{t-1}(i) \cdot \mathrm{rad}_{x_0}(X_0) \\
&\leq c \log^{(t)} i \cdot \mathrm{rad}_{x_0}(X) + \epsilon \cdot c \log^{(t)} i \cdot \mathrm{rad}_{x_0}(X) - c \cdot \mathrm{rad}_{x_0}(X_0)/(10\varphi_{t-1}(i)) \\
&\leq c \log^{(t)} i \cdot \mathrm{rad}_{x_0}(X) + \mathrm{rad}_{x_0}(X) \cdot \log^{(t)} i/(170\varphi_t(i)) - \mathrm{rad}_{x_0}(X)/(160\varphi_{t-1}(i)) \\
&\leq c \log^{(t)} i \cdot \mathrm{rad}_{x_0}(X),
\end{aligned}$$

the third inequality using Claim 10. In a similar manner, it can shown that the same holds for $t = 1$.

□

*Proof of Theorem 1.* Take $c = 2^{18} \log^{(t)} n$ (recall that $t = (\log^* n)/2$ and indeed $t \leq \log^* c$). Note that $1/\epsilon = O(c\varphi_t(n))$ and the parameter $k = O(c\log(1/\epsilon)) = O(c \log\log\log n)$. The increase in radius is $\mathrm{rad}(T) \leq O(c^2 \mathrm{rad}(X))$, and plugging in these parameters to Lemma 7 implies that the expected stretch for any edge $(u, v) \in E$ is bounded by

$$\begin{aligned}
\mathbb{E}_{T\sim\mathcal{T}}[d_T(u, v)] &\leq O\left(c^2 \cdot \log n \cdot k/\epsilon\right) \\
&= O(c^4 \log n \cdot \log\log\log n \cdot \varphi_t(n)) \\
&= O\left(\log^{(1)} n \cdot \log^{(2)} n \cdot \log^{(3)} n \cdots \log^{((\log^* n)/2)} n \cdot \left(\log^{((\log^* n)/2)} n\right)^4 \cdot \log^{(3)} n\right)
\end{aligned}$$

□

# C Proof of some claims

*Proof of Claim 10.* We prove by induction on $t$. The base case where $t = 2$ holds since $\log\log(i^a) = \log(a \log i) = \log\log i + \log a$. Assume the claim holds for $t$ and we prove for $t+1$

$$\begin{aligned}
\log^{(t+1)}(i^a) &= \log\left(\log^{(t)}(i^a)\right) \\
&\leq \log\left(\log^{(t)} i + (\log a)/\varphi_{t-1}(i)\right) \\
&= \log\left(\log^{(t)} i \cdot \left(1 + (\log a)/(\varphi_{t-1}(i) \cdot \log^{(t)} i)\right)\right) \\
&\leq \log\left(\log^{(t)} i \cdot \left(2^{\log a/\varphi_t(i)}\right)\right) \\
&= \log^{(t+1)} i + (\log a)/\varphi_t(i).
\end{aligned}$$

The first inequality uses the induction hypothesis, the last inequality holds because $\log a \leq 0$, and $1 + x \leq 2^x$ for $x \leq 0$. □

**Claim 13.** *For any* $c \geq 0$, $b \geq 2$ *and* $0 \leq t \leq \log^* b$

$$\log^{(t)}(b + c) \leq (\log^{(t)} b) + c.$$

*Proof.* By induction on $t$, for $t = 0$ it holds since by definition $\log^{(0)}(b + c) = b + c$. Assume for $t - 1$ and prove for $t$

$$\begin{aligned}
\log^{(t)}(b + c) &= \log^{(t-1)}(\log(b \cdot (1 + c/b))) \\
&\leq \log^{(t-1)}(\log b + (c \log e)/b) \\
&\leq \log^{(t)} b + (c \log e)/b \\
&\leq \log^{(t)} b + c.
\end{aligned}$$

We used the induction hypothesis in the second inequality. □

*Proof of Claim 11.*

$$\begin{aligned}
\log^{(t)}(ai) &= \log^{(t-1)}(\log i + \log a) \\
&= \log^{(t-1)}(\log i \cdot (1 + (\log a)/\log i)) \\
&\leq \log^{(t-1)}\left(\log i \cdot e^{(\log a)/\log i}\right) \\
&= \log^{(t-2)}(\log\log i + (\log a \cdot \log e)/\log i) \\
&\leq \log^{(t)} i + (2 \log a)/\log i.
\end{aligned}$$

The last inequality we use Claim 13 with $b = \log\log i > \log e$. □